# Time-dependent phase quantification and local structure analysis of hydroxide-activated slag via X-ray total scattering and molecular modeling

Kai Gong[1,2], Claire E. White[1,*]

[1]Department of Civil and Environmental Engineering and the Andlinger Center for Energy and the Environment, Princeton University, Princeton, New Jersey 08544, United States

[2]Department of Materials Science and Engineering, Massachusetts Institute of Technology, Cambridge, MA 02139, United States (current address)

* Corresponding author: Phone: +1 609 258 6263, Fax: +1 609 258 2799, Email: whitece@princeton.edu

Postal address: Department of Civil and Environmental Engineering, Princeton University, Princeton NJ 08544, USA

**Abstract**

Here, an approach to quantify the amorphous-to-disordered/crystalline transformation occurring in NaOH-activated ground granulated blast-furnace slag (GGBS) is outlined that combines atomistic modeling with *in situ* pair distribution function (PDF) analysis. Firstly, by using force-field molecular dynamics (MD) simulations, a detailed structural representation is generated for the amorphous GGBS that is in agreement with experimental X-ray scattering data. Use of this structural representation along with literature-derived structures for the reaction products allows for real space X-ray PDF refinement of the alkaline activation of GGBS, resulting in the quantification of all phases and the degree of reaction (DOR) as a function of reaction time. All phases and the DOR are seen to approximately follow a logarithmic-type time-dependent behavior up to 5 months, while at early age (up to 11 hours) the DOR is accurately captured by a modified pseudo-single step first-order reaction model. Lastly, the evolution of DOR is found to agree with several other complementary *in situ* data containing quantitative reaction information, including isothermal conduction calorimetry, Fourier transform infrared spectroscopy, and quasi-elastic neutron scattering.

# 1 Introduction

The study of chemical reactions involving dissolution and/or formation of amorphous phases is crucial to many industrially important applications, including bioglass dissolution, alkali-activated materials (AAMs), blended ordinary Portland cement (OPC), glass corrosion, and nuclear waste encapsulation [1-4]. For AAMs, the associated chemical reactions are particularly challenging to investigate, given the amorphous or disordered nature of both the precursor materials and the resulting alkali-activated binder. Hence, methods that are able to accurately quantify the amorphous phase(s) together with the overall degree of reaction (DOR) are extremely valuable, particularly since this quantitative information can be utilized to further optimize these chemical reactions to obtain desired material properties and performance.

In the field of cement research, one popular method used to quantify an amorphous phase in an amorphous-crystalline mixture is the combination of Rietveld refinement with the Partial Or No Known Crystal Structure (PONKCS) approach based on X-ray (or neutron) diffraction, which has been successfully applied to AAMs [5, 6], blended OPC systems [7-10] and other types of cementitious materials [11]. This PONKCS method often requires the use of an internal or external standard, although an intensity-based direct decomposition method (also based on XRD data) that does not require any standards for quantification of the amorphous phase has also been developed and applied to blended OPC cements [12] and AAMs [13]. Furthermore, different experimental methods have been used in the cement literature for direct or indirect estimation of the overall DOR for AAMs, blended OPC systems and other types of cements. These include deconvolution of $^{29}$Si and/or $^{27}$Al nuclear magnetic resonance (NMR) spectra [10, 14-16], image analysis based on backscattered scanning electron microscopy (SEM) [8, 17-19], selective acid or alkaline dissolution [8, 9, 19-21], differential scanning calorimetry (DSC) [8], non-evaporable water or portlandite content from thermogravimetric analysis (TGA) [9, 10, 18, 22, 23], isothermal conduction calorimetry (ICC) [24-26], and chemical shrinkage [8], with their relative strengths and shortcomings discussed in refs. [8, 19, 24, 27]. A summary of the suitability of these different methods for the estimation of DOR for different types of SCMs in blended OPC systems has been presented in ref. [24]. This reference states that the XRD-based method is generally reliable and considered a promising technique due to ease of access to XRD instruments, relatively rapid data acquisition, and additional information contained in the data on reaction products.

The XRD-based phase quantification method can be further enhanced by exploiting synchrotron-based X-rays, that allow high-resolution XRD data to be collected in a matter of minutes (or even seconds). As a result, the phase transformations that occur during different chemical processes (including the formation of AAMs and OPC systems) can be probed *in situ*, which is impossible for many of the above-mentioned methods (e.g., selective dissolution, DSC, TGA and SEM). The use of high energy and high brilliance synchrotron-based X-rays also allows for the *in situ* collection of X-ray total scattering data, which, when Fourier transformed from reciprocal to real space gives the X-ray pair distribution function (PDF) [28], which is ideal for studying the local atomic structure of amorphous and disordered materials [12]. In recent years, *in situ* PDF analysis has emerged as an important characterization technique for investigating the structural evolution of cement-based materials, including reaction kinetics [29-32], carbonation mechanisms [3, 33], drying shrinkage [34], and atomic changes under loading [35, 36].

In this investigation, we utilize *in situ* synchrotron-based XRD and subsequent X-ray PDF analysis to track the phase transformations and associated evolution of the local atomic structure occurring during sodium hydroxide activation of amorphous GGBS. According to previous investigations [5, 13, 17, 18, 37-39], the main reaction product in sodium hydroxide-activated GGBS is a sodium-containing calcium-alumino-silicate-hydrate (C-(N)-A-S-H) gel with a structure resembling that of poorly ordered C-S-H (I) [40]. In addition to the main C-(N)-A-S-H binder gel, secondary layered double hydroxide (LDH) phases are often observed in alkali-activated GGBSs [37, 38, 41-43], along with the presence of a substantial amount of unreacted GGBS even at an advanced age of curing (e.g., 6-12 months) [14, 15, 17, 18]. Essentially, this alkaline activation process converts the amorphous GGBS to disordered/crystalline reaction products. Quantification of such amorphous-to-disordered/crystalline transformation processes is critical if we are to fully determine the reaction kinetics and mechanisms occurring in alkali-activated GGBSs, other types of AAMs as well as blended cements.

A major goal of this investigation is to quantify such an amorphous-to-disordered/crystalline transformation occurring during hydroxide activation of amorphous GGBS, based on a high temporal resolution quantification method that combines atomistic modeling with *in situ* PDF analysis. Specifically, we first obtained structural representations of all identified phases (based on the synchrotron-based XRD data) from the literature, apart from amorphous GGBS where the structural representation is generated using force-field-based molecular dynamics (MD) simulations and validated using experimental X-ray PDF data. The resulting amorphous structural representation for the GGBS along with structural representations for the different reaction products (i.e., the C-(N)-A-S-H gel and secondary LDH phases) are used to simulate the X-ray PDF curve of each phase, which are then collectively refined against the experimental PDF data using a least squares refinement method. This refinement process enables the calculation of the relative percentage of each phase in the NaOH-activated GGBS binder as a function of reaction time, including the amorphous unreacted GGBS, along with the overall DOR. Analysis of these time-dependent data provides insight on the reaction kinetics of individual phases. Lastly, to validate the quantification method, the evolution of DOR from PDF analysis is compared with *in situ* reaction data obtained using three other experimental techniques, specifically ICC and Fourier transform infrared spectroscopy (FTIR) data collected on the same NaOH-activated GGBS, as well as *in situ* quasi-elastic neutron scattering (QENS) data collected on a similar hydroxide-activated GGBS [44]. This investigation highlights the power of using real space PDF refinement in combination with atomistic modeling to quantify amorphous-to-disordered/crystalline (and amorphous-to-amorphous) transformations that are ubiquitous in many important chemical processes and materials applications, including AAMs.

## 2 Materials & Methods

### 2.1 Materials & sample preparation

The GGBS used in this investigation had an oxide composition of 42.5 wt. % CaO, 34.5 wt. % $SiO_2$, 11.7 wt. % $Al_2O_3$, and 7.3 wt. % of MgO (along with other minor oxides), based on X-ray fluorescence (XRF) measurements [3]. For GGBS activation, a sodium hydroxide solution was prepared using NaOH pellets (Sigma-Aldrich, reagent grade) and deionized water. The mix proportions were 50 g of $H_2O$ and 4 g of $Na_2O$ for every 100 g of GGBS. Firstly, the activating solution was prepared by dissolving the NaOH pellets in deionized water, after which it was left to cool down to ambient temperature. The mixture was then hand-mixed for 2 mins immediately before being subjected to the various measurements as detailed below. For samples that were measured at later ages (e.g., 5 months), they were sealed in airtight containers after mixing and left to cure at ambient conditions.

## 2.2 Experimental details

### 2.2.1 High-resolution XRD and X-ray total scattering

Immediately after mixing, high-resolution XRD and X-ray total scattering measurements were performed at ambient temperature, using the 11-ID-B beamline at the Advanced Photon Source, Argonne National Laboratory. Both types of data (high-resolution XRD and X-ray total scattering) were collected simultaneously on the mixture *in situ* for approximately 11 hours using a Perkin-Elmer amorphous silicon 2D image plate detector [45] and a wavelength of 0.2113 Å. The XRD data were collected at a sample-to-detector distance of ~950 mm, while the X-ray total scattering data were collected at a distance of ~180 mm. Subsequent measurements were performed on the same sample at 15, 24, 55 and 77 hours using the same protocol, with a final measurement collected at 5 months. The Fit2D program [46, 47] was employed to convert data from 2D to 1D, with $CeO_2$ as the calibration sample. The PDF ($G(r)$) was then calculated by taking a sine Fourier transform of the measured total scattering function $S(Q)$ (similar to our previous investigations [37, 38]), as illustrated in Equation (1) [28]:

$$G(r) = \frac{2}{\pi} \int_{Q_{min}}^{Q_{max}} Q[S(Q) - 1] \cdot \sin(Qr) \, dQ \quad (1)$$

where $Q$ is momentum transfer given in Equation (2), and $r$ is the interatomic distance.

$$Q = \frac{4\pi \cdot \sin\theta}{\lambda} \quad (2)$$

In Equation (2), $\theta$ and $\lambda$ are the scattering angle and wavelength of the monochromatic X-rays, respectively. PDFs were obtained using PDFgetX3 [48] with a $Q_{max}$ of 20 Å$^{-1}$, following standard data reduction procedures. The instrument parameters were refined using a standard calibration sample (nickel) and the refinement program PDFgui [49]. The refined parameters ($Q_{broad}$ = 0.00086 Å$^{-1}$ and $Q_{damp}$ = 0.03787 Å$^{-1}$) were used to calculate the simulated PDFs of individual phases based on their atomic structures using the PDFgui program [49].

### 2.2.2 Fourier transform infrared spectroscopy (FTIR)

Attenuated total reflectance (ATR)-FTIR measurements were performed on neat GGBS, and the NaOH-activated GGBS mixture at different curing times (sealed curing was carried out in a plastic container) over a period of 8 days. A PerkinElmer FTIR instrument (Frontier MIR with a Frontier UATR diamond/ZnSe attachment) with an $N_2$ flow was used. 32 scans were collected from 600 to 4000 cm$^{-1}$ with a resolution of 4 cm$^{-1}$ for each measurement.

*2.2.3 Isothermal conduction calorimetry (ICC)*

ICC measurement was performed on the same NaOH-activated GGBS immediately after mixing using a TAM Air isothermal calorimeter (TA Instruments), following a similar procedure adopted in our previous investigation [44]. Heat release was measured at 25 °C over seven days for 5 g of NaOH-activated GGBS, using 5 g of deionized water as the reference.

## 2.3 Computational details

A detailed atomic structural representation for the GGBS was generated using force-field MD simulations following a similar procedure adopted in our previous investigation [50]. Specifically, a cubic unit cell consisting of ~2000 atoms with a chemical composition of $(CaO)_{363}(MgO)_{87}(Al_2O_3)_{55}(SiO_2)_{275}$ (similar to the main $CaO$-$MgO$-$Al_2O_3$-$SiO_2$ composition of the GGBS used in the experiments) was first equilibrated at a temperature of 5000 K for 1 ns to ensure the loss of the memory of the initial structure and to reach a "melt" state. The structure was then quenched from 5000 to 2000 K at a rate of 1 K/ps followed by 1 ns of equilibration at 2000 K. The equilibrated structure was further quenched from 2000 to 300 K in 3 ns, followed by another 1 ns of equilibration at 300 K. The canonical *NVT* ensemble (with the Nosé Hoover thermostat) and a time step of 1 fs were used for all the MD simulation steps outlined above, while the density of the unit cell volume was adjusted to numerically estimated values (calculation method [51] and details are given in Section 1 of the Supplementary Material), as discussed in our previous investigation [50]. The interatomic potential developed by Guillot for crystals and melts covering the $CaO$-$MgO$-$Al_2O_3$-$SiO_2$ system [52] were employed for the MD simulations, as described by Equation (3).

$$U_{ij}(r_{ij}) = \frac{z_i z_j}{r_{ij}} + B_{ij} e^{-\frac{r_{ij}}{\rho_{ij}}} - \frac{C_{ij}}{r_{ij}^6} \tag{3}$$

where $z_i$ and $z_j$ are the effective charge associated with atom $i$ and $j$ respectively, $r_{ij}$ is the interatomic distance between atom pair *i-j*, and $\rho_{ij}$, $B_{ij}$ and $C_{ij}$ are the energy parameters obtained by refinement against experimental data [52]. The force-field parameters adopted are given in Table 1. All the simulations were carried out using the QuantumATK software package [53, 54]. As shown in ref. [50], the above MD simulation protocol is able to generate amorphous structural representations that are representative of GGBS local atomic structures, as determined by comparison of simulated and experimental PDF datasets.

Table 1. Summary of the force-field parameters [52] used for the MD simulations.

| Atom | $z$ (e) | Metal-oxygen pair | $B$ (eV) | $1/\rho$ (Å$^{-1}$) | $C$ (Å$^6$ eV) | $r_{cut}$ (Å) for the three terms in Equation (3) | |
|---|---|---|---|---|---|---|---|
| | | | | | | 1$^{st}$ term | 2$^{nd}$ & 3$^{rd}$ terms |
| Si | 1.89 | Si-O | 50306.2 | 6.21118 | 46.2979 | 12 | 7.5 |
| Al | 1.4175 | Al-O | 28538.5 | 5.81395 | 34.5779 | 12 | 7.5 |
| Mg | 0.945 | Mg-O | 32652.7 | 5.61798 | 27.281 | 12 | 7.5 |
| Ca | 0.945 | Ca-O | 155668 | 5.61798 | 42.2598 | 12 | 7.5 |
| O | -0.945 | O-O | 9022.82 | 3.77359 | 85.0924 | 12 | 7.5 |

## 2.4 Method of quantification using PDF data

The detailed atomic structural representation for amorphous GGBS, along with the structural representations of the reaction products (identified using XRD) obtained from the literature, were used to simulate the PDF contributions from each phase, which were subsequently refined against the experimental PDF data for NaOH-activated GGBS over an $r$ range of 1-15 Å, using the least squares refinement method implemented in the PDFgui program [49]. The upper limit of 15 Å (= 30/2) was selected to avoid the finite boundary effects of the GGBS structural representation, which has a cell size of ~30×30×30 Å$^3$. The only parameters that were refined for the phases in PDFgui were the scale factor (a reflection of the relative phase quantity) and 'delta2' (accounts for the peak sharpening at low $r$ distance due to correlated thermal motions [55]). The isotropic atomic displacement parameters were set at $u_{ii}$ = 0.003 Å$^2$ for all phases. The lattice parameters of all the reaction product phases were refined against the 5-month X-ray PDF data and then kept constant for the refinement of all other datasets. Similarly, the lattice parameters of the GGBS representation were refined against the X-ray PDF data of the neat GGBS, and then kept constant for all other refinements. Refinement of all NaOH-activated GGBS PDF datasets led to time-dependent scale factors for each individual phase. These scale factors were analyzed to assess (i) the evolution of the various phases that may be present during the alkaline activation reaction and (ii) the DOR.

## 3  Results & Discussion

### 3.1  Characterization and modeling of the neat GGBS

The high-resolution XRD pattern for the neat GGBS is presented in Figure 1a, which shows that this GGBS is dominated by a broad diffuse hump centered at a $Q$ value of ~2.1 Å$^{-1}$ attributed to an amorphous phase, known to be predominantly a Ca- and Mg-containing aluminosilicate (CMAS) glass [37]. The XRD data also reveal the presence of several minor crystalline phases in the neat GGBS, including calcite, vaterite, aragonite and åkermanite. The corresponding X-ray PDF data is presented in Figure 1b, which reveals the absence of any obvious long-range ordering above ~12 Å and as such indicates that this GGBS is largely amorphous with the minor crystalline phase quantities being minimal. Despite the lack of obvious long-range ordering, the GGBS clearly exhibits short- (< ~3 Å) and medium-range (~3-10 Å) ordering that is typical of amorphous aluminosilicate materials [30, 37, 50, 56], as seen in the inset of Figure 1b. Assignment of the nearest atom-atom correlations below ~3 Å for the neat GGBS or other amorphous aluminosilicate materials is relatively straightforward (as shown in the inset of Figure 1b). Based on the existing literature data [37, 50, 57-59], the peaks at ~1.62, ~1.98, ~2.34 and ~2.66 Å are assigned to Si/Al$^{IV}$-O, Mg-O, Ca-O and O-O correlations, respectively. However, the atom-atom correlations above ~3 Å are challenging to assign without an accurate structural representation due to the coincidence of individual correlations [50, 57, 59].

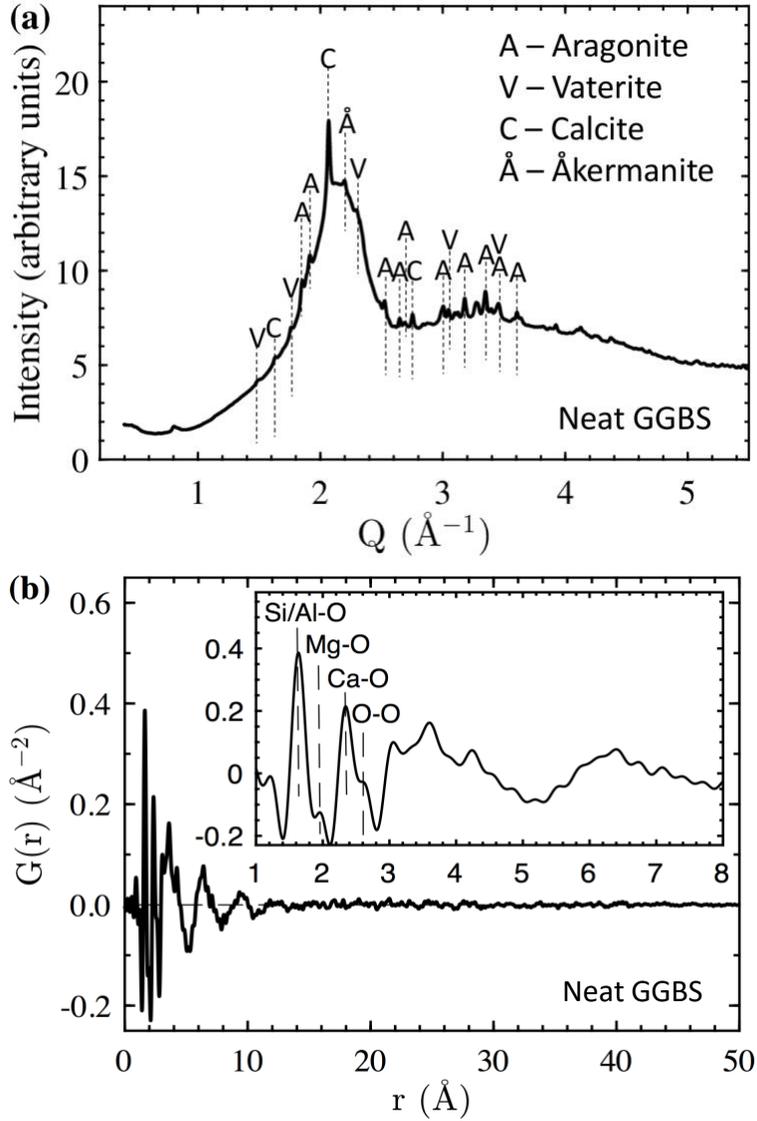

Figure 1. (a) High-resolution XRD and (b) X-ray PDF for the neat GGBS. The inset figure in (b) is a zoom from 1 to 8 Å.

A detailed structural representation for the GGBS has been generated with the melt-and-quench approach using force-field MD simulations (as outlined in Section 2.3). The resulting representation (see Figure 2a) provides good agreement with the corresponding experimental X-ray PDF data as shown in Figure 2b, where the short- (< 3 Å), medium- (~3-10 Å), and long-range (> 10 Å) ordering are mostly captured by the structural representation. The partial X-ray PDFs calculated using the structural representation in Figure 2a are shown in Figure 2b (calculation details are shown in Section 2 of the Supplementary Material), which reveal that the medium-range ordering between ~4 and 5 Å is mainly attributed to the second nearest Si-O and Ca-O correlations in the GGBS, whereas the medium-range ordering between ~5 and 8 Å is mainly due to the third nearest Ca-O correlation and the second nearest Ca-Ca and Ca-Si correlations. The double peak at ~3.1 and ~3.6 Å is predominantly assigned to the nearest Ca-Si correlations in edge-sharing and corner-sharing configurations, respectively (as has been explicitly shown in our previous investigation [50]), with some contributions from the nearest Ca-Al and Ca-Ca correlations. These assignments are consistent with our previous findings for a neat GGBS with slightly different chemical composition [50]. Furthermore, the nearest interatomic distances (~1.62, ~1.75, ~2.03, ~2.42 Å for Si-O, Al-O, Mg-O and Ca-O correlations, respectively) and the average coordination numbers of Si, Al, Mg and Ca atoms (4.00, 4.01, 5.20 and 6.73, respectively) agree reasonably well with those from experiments (including our X-ray PDF data in Figure 1b for interatomic distances) and simulations on GGBS and aluminosilicate glasses [50, 57-62].

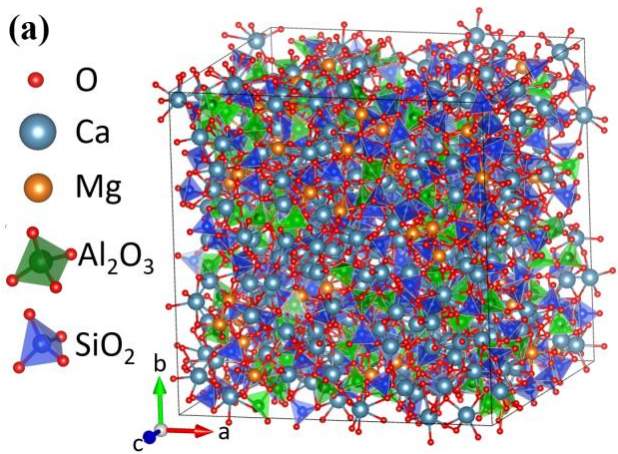

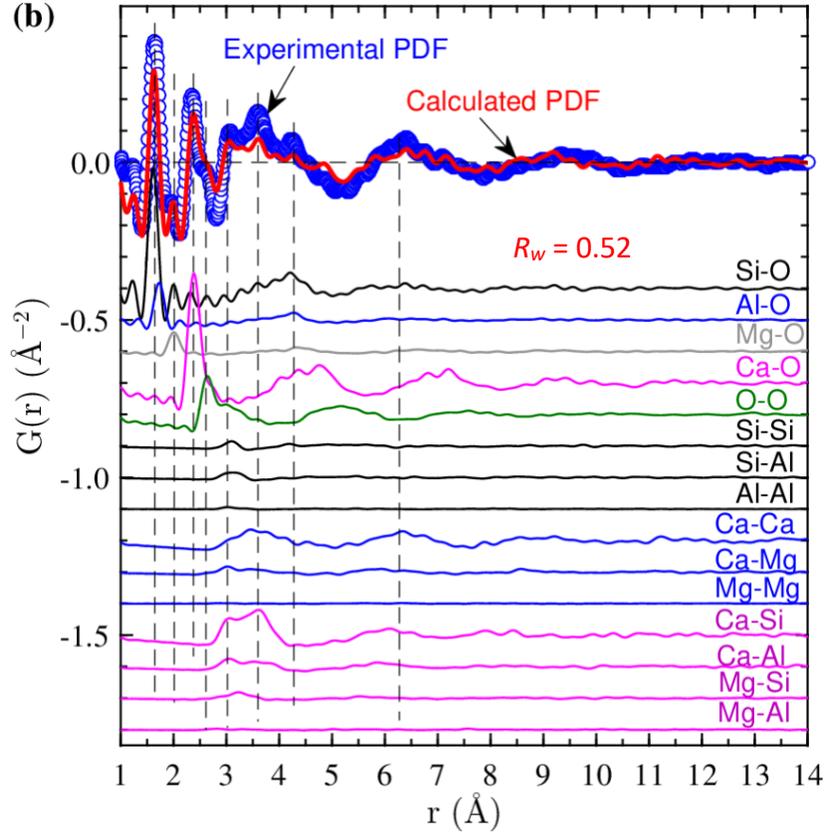

Figure 2. (a) Structural representation for the GGBS obtained using force-field MD simulations (box size: ~30×30×30 Å$^3$). (b) Comparison between the experimental X-ray PDF and the simulated X-ray PDF calculated using the structural representation in (a) along with the level of agreement ($R_w$ value). The $R_w$ value in (b) denotes the extent of agreement between the experimental and simulated PDF data, with the calculation details of $R_w$ shown in Section 2 of the Supplementary Material. Partial X-ray PDFs for the different atom-atom correlations present in the structural representation are also shown in (b).

### 3.2 Characterization and modeling of the final NaOH-activated GGBS binder

Figure 3a shows a high-resolution XRD pattern of the NaOH-activated GGBS binder that has been cured for 5 months. Phase identification confirms, as expected, that a calcium silicate hydrate-type gel phase (i.e., C-(N)-A-S-H) is the main reaction product formed in the sample, where the main peaks centered at $Q$ values of 0.50, 1.20, 2.05, 2.25 and 3.43 Å$^{-1}$ are assigned to 002, 101, 020/$\bar{2}$20, 200 and $\bar{2}$40 reflections, respectively [63]. The presence of these peaks is generally consistent with previous investigations on hydroxide-activated GGBSs [5, 13, 17, 18, 37-39] and synthetic C-S-H-type gels [40, 63-68]. The 002 reflection at 0.50 Å$^{-1}$ corresponds to a basal spacing of ~12.6 Å between the calcium oxide layers in the C-(N)-A-S-H gel structure, which agrees well with that of poorly ordered C-S-H (I) [40] and synthetic C-(N)-A-S-H gels [66, 69], although the basal spacing for synthetic C-S-H-type gels has been seen to vary significantly between ~9 and ~16 Å, depending on Ca/Si ratio, Al/Si ratio, alkali content, interlayer water content, and presence of cross-linking [64, 67-71].

In addition to the main binder gel (i.e., C-(N)-A-S-H), two LDH phases have been identified in the 5-month sample: an AFm-group calcium hemicarboaluminate hydrate phase ($C_4A\underline{c}_{0.5}H_{12}$ [41], abbreviated as Hc) and a hydrotalcite-like phase (e.g., $Mg_6Al_2(OH)_{16}CO_3 \cdot 4.5H_2O$ [42], abbreviated as Ht). The Ht phase is commonly observed in hydroxide-activated GGBS systems [37, 38], whereas the formation of the Hc phase is not reported as frequently in alkali-activated systems. One possible reason for the clear identification of the Hc phase in Figure 3a is the use of high-resolution synchrotron-based XRD in this investigation that allows for differentiation of the two LDH phases. Nevertheless, Hc phase has been observed in $Na_2CO_3$-activated GGBSs [72] and blended Portland cements [73, 74], and an AFm-type phase ($C_4AH_{13}$) has been observed in NaOH-activated GGBS [43]. The corresponding X-ray PDF data in Figure 3b show that the 5-month NaOH-activated GGBS binder possesses distinct atom-atom correlations up to ~40 Å, beyond which the PDF is seen to be structureless. This is consistent with previous PDF analysis on the structure of synthetic C-S-H gels where atomic ordering out to ~40 Å has been observed [75, 76].

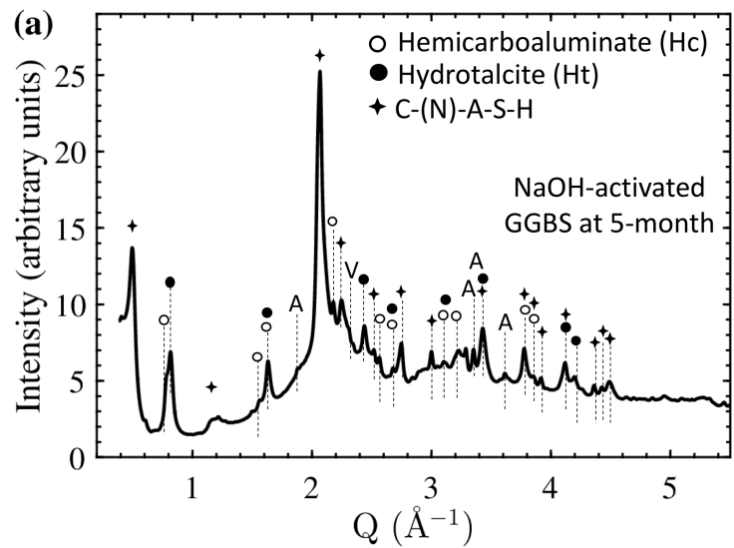

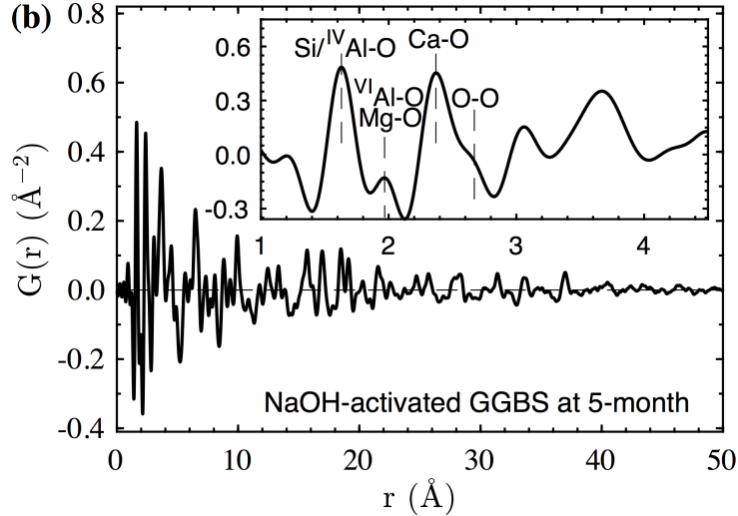

Figure 1. (a) High-resolution XRD and (b) X-ray PDF for the NaOH-activated GGBS that has been cured for 5 months. Identification of the Ht and Hc phases and C-(N)-A-S-H gel in (a) are based on ICSD #107626 [42], #263124 [41], and # 152489 [77], respectively. The inset figure in (b) is a zoom of the PDF below 4.5 Å.

A broad diffuse region at a $Q$ value of ~2.1 Å$^{-1}$ is clearly visible in the XRD pattern in Figure 3a, indicating the presence of a considerable amount of unreacted GGBS in the 5-month sample, as commonly reported for alkali-activated GGBS binders at a similar age [14, 17, 37]. With the generation of a detailed structural representation for the amorphous GGBS (Figure 2a), it is now possible to quantify the unreacted amorphous phase fraction along with the three reaction products using the method outlined in Section 2.4. The results for the 5-month sample using this refinement process are presented in Figure 4a, where it is clear that the total simulated PDF (including contributions from the four different phases, i.e., unreacted GGBS, C-(N)-A-S-H gel, Ht and Hc) gives relatively good agreement with the experimental data. Analysis of the data involving possible contributions from calcium carbonate phases (e.g., calcite, aragonite and vaterite) was also carried out; however, the contributions from these carbonate phases were found to be relatively small and did not change the goodness of fit ($R_w$ value), so we restrict our analysis to the four main phases. The scale factors shown in Figure 4a reflect the relative quantities of the different phases present in the sample. From the scale factor and the simulated X-ray PDF of each phase (Figure 4a), it is clear that the X-ray PDF of the 5-month sample is dominated by atom-atom correlations from the C-(N)-A-S-H gel, especially for the medium- (~3-10 Å) and long-range (> 10 Å) ordering. For the short-range ordering, the unreacted GGBS phase also contributes substantially to the experimental data, especially for the nearest Si/Al$^{IV}$-O and Ca-O correlations located at ~1.62 and ~2.37 Å, respectively.

The structural representation for the C-(N)-A-S-H gel phase used in Figure 4a is a 14 Å tobermorite structure that has been generated using DFT calculations [78], and has a composition of Ca$_9$Si$_{11}$Al$_1$Na$_1$O$_{50}$H$_{34}$. Based on this structural representation, the partial X-ray PDFs for the different atom-atom pairs have been calculated and are shown in Figure 4b. These partial X-ray PDFs together with those from neat GGBS (Figure 2b) can be used to further understand the experimental PDF data of the 5-month NaOH-activated GGBS. For example, from Figure 4b, we can see that the nearest Ca-Si correlation in the C-(N)-A-S-H gel exhibits a bimodal feature with individual peaks centered at ~3.04 and ~3.67 Å, similar to the neat GGBS (see Figure 2b). Previously, the PDF peak at ~3.67 Å in alkali-activated GGBS and C-S-H-type gels has been assigned mainly to Ca-Si or Ca-T (T = tetrahedral Si and Al) correlations [31, 34, 37, 38, 68]; however, the partial PDFs in Figure 4b and Figure 2b suggest that this peak also has large contributions from the nearest Ca-Ca correlation and the second nearest Si-O correlation in the C-(N)-A-S-H gel structure, as well as contributions from the unreacted GGBS for the case of alkali-activated GGBS binder. The assignment of the peak at ~3.67 Å to Ca-Si, Ca-Ca and Si-O correlations is consistent with previous X-ray PDF investigations on synthetic C-S-H gels [65, 79]. The peak at ~3.04 Å in NaOH-activated GGBS (Figure 4a) consists of contributions from C-(N)-A-S-H gel (specifically Ca-Si, Si-T, as seen in Figure 4b), unreacted GGBS (Ca-T, Si-T and O-O, as seen in Figure 2b) as well as the Ht phase (Figure 4a).

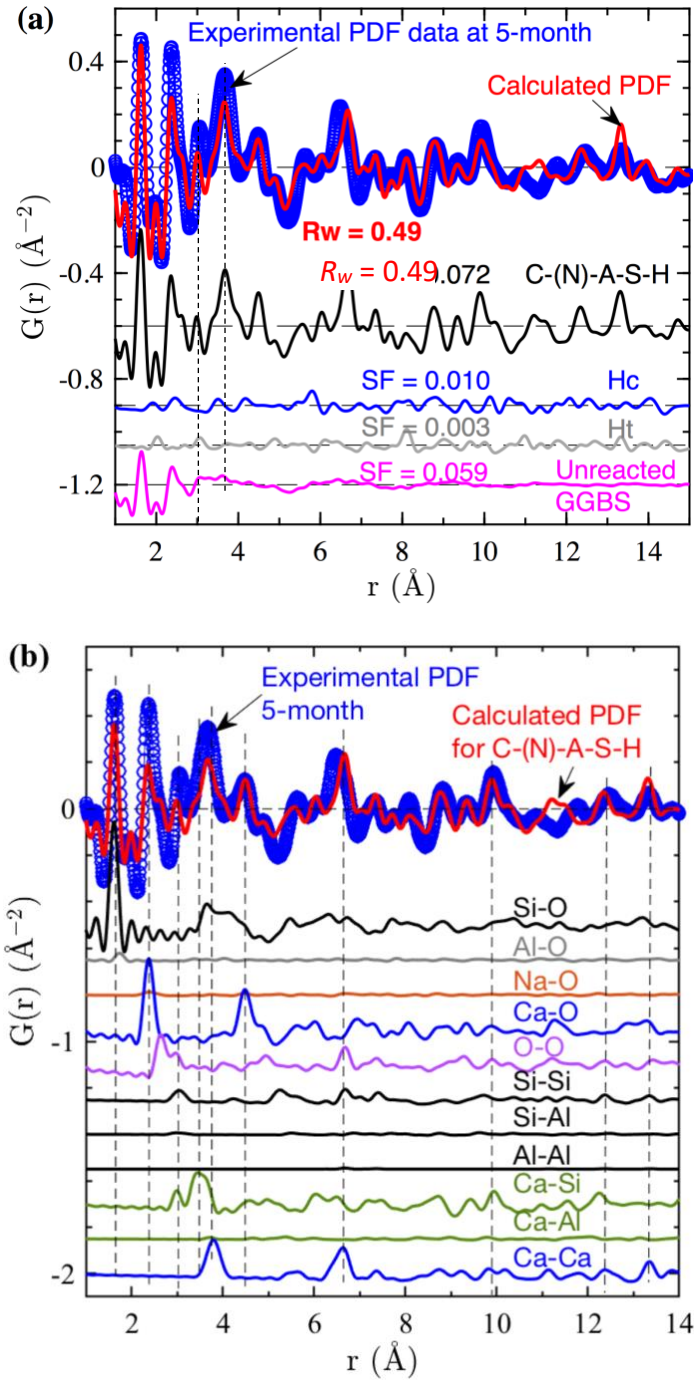

Figure 4. (a) Comparison of the experimental X-ray PDF data for the 5-month NaOH-activated GGBS sample and the calculated X-ray PDF based on three structural representations available in the literature (i.e., ICSD #107626 [42] for Ht, ICSD #263124 [41] for Hc, and C-A-S-H[Na,2H] in ref. [78] for the C-(N)-A-S-H gel) together with the GGBS structural representation presented in Figure 2. The PDF contribution from each phase is also shown in (a) along with their respective scale factors (SF). (b) The experimental X-ray PDF data and the simulated PDF contribution from the C-(N)-A-S-H gel obtained from the multiphase refinement process, along with the partial X-ray PDFs for different atom-atom pairs in the C-(N)-A-S-H gel. See Section 2 of the Supplementary Material for details on the calculation of the partial X-ray PDFs.

Figure 5a compares the experimental PDFs for the neat GGBS and the 5-month NaOH-activated GGBS binder, which shows that obvious structural changes (both short- and long-range) have occurred due to alkaline activation of GGBS. For instance, the PDF peak at ~3.60 Å is seen to shift to ~3.67 Å and experience a ~70% increase in the peak intensity after alkaline activation. This change invoked by alkaline activation can be attributed mainly to two major differences between the partial PDFs of C-(N)-A-S-H gel and neat GGBS, as seen in Figure 5b-c. The first difference involves the Ca-Si correlation, where, as seen in Figure 5b, this correlation for C-(N)-A-S-H gel has a much higher intensity at ~3.60 Å than that of the neat GGBS. The second difference involves the Ca-Ca correlation, which is also more intense in C-(N)-A-S-H gel than neat GGBS and positioned at a larger distance (i.e., ~3.79 Å for the gel and ~3.42 for GGBS, see Figure 5c). This larger Ca-Ca distance for the C-(N)-A-S-H gel structure can be attributed to (i) the larger Ca-O bond length (Figure 5a) and (ii) a narrower Ca-O-Ca angular distribution (see Figure S1 in Supplementary Material) for the C-(N)-A-S-H gel structural representation as compared to the structural representation for neat GGBS.

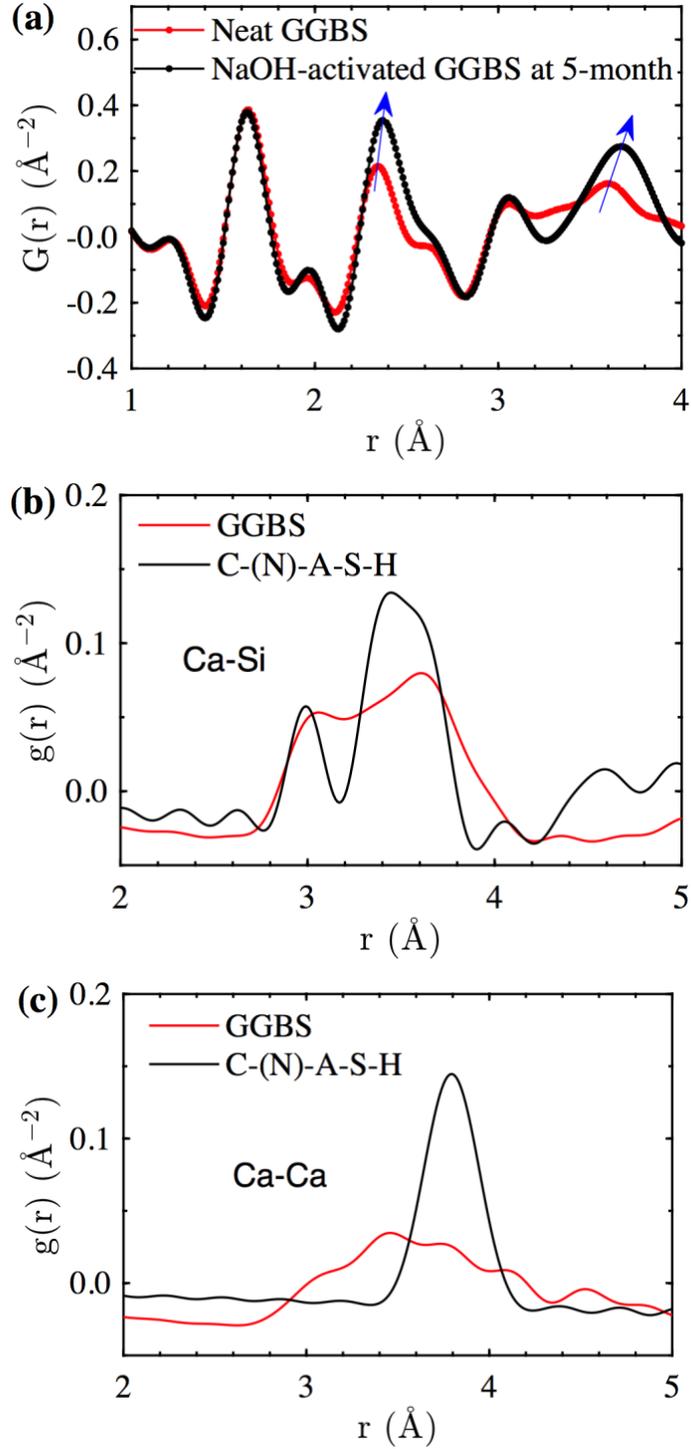

Figure 5. Comparison of (a) the experimental X-ray PDFs of neat GGBS and 5-month NaOH-activated GGBS, and (b) the simulated Ca-Si and (c) Ca-Ca partial PDFs obtained from the structural representations of neat GGBS (Figure 2b) and C-(N)-A-S-H gel (Figure 4b).

Accurate assignment of the atom-atom correlations contained within the PDF data is important since this information can reveal new insight on the local structure of the material and, for *in situ* measurements, time- or environment-dependent changes that may be linked to underlying chemical mechanisms. In this study the *in situ* measurement of the alkali-activation reaction involving GGBS is of interest, with the aim to uncover new insight on how the alkaline activation reaction proceeds (as will be discussed in the next section). However, it is clear from Figures 2b and 4a that there are obvious discrepancies between the experimental PDF data and the simulated PDFs based on structural representations, suggesting that these structural representations can be further improved, especially for C-(N)-A-S-H gel since the current structural representation is crystalline with a relatively small unit cell. From experiments it is known that C-(N)-A-S-H gel is disordered/nanocrystalline with varying stoichiometry. Other limitations associated with the current structural representation for C-(N)-A-S-H are that the silicate chains are infinite in length, and all Al bridging sites are on one side of the interlayer space, which is likely to differ from reality. Nevertheless, as will be shown below, we can still use these structural representations to robustly quantify the time-dependent behavior of alkali-activated GGBS and specifically the evolution of the individual phases.

### 3.3 Phase evolution during alkaline activation of GGBS

In this section, *in situ* high-resolution XRD data are analyzed to elucidate the phase evolution behavior that occurs during hydroxide activation of the GGBS. As illustrated in Figure 6a, the characteristic Bragg peaks denoting the reaction products (i.e., XRD peaks located at $Q$ values of 0.47-0.50, 1.20, 2.05, 2.25 and 3.43 $\text{Å}^{-1}$ assigned to the C-(N)-A-S-H gel, 0.82, 1.63 and 2.43 $\text{Å}^{-1}$ assigned to the Ht phase [42], and 0.78, 1.55, 1.63 and 2.17 $\text{Å}^{-1}$ assigned to the Hc phase [41]) increase in intensity with the progress of reaction, whereas the broad diffuse region (centered at a $Q$ value of ~2.1 $\text{Å}^{-1}$) representing unreacted amorphous GGBS diminishes with time. A closer examination of peaks assigned to the two LDH phases at ~0.78 and 0.82 $\text{Å}^{-1}$ in Figure 6b (which correspond to the basal spacing for the Hc and Ht phases, respectively) reveals that both LDH phases already start to form after 0.2 hours of reaction and the formation of the Hc phase appears to be more pronounced than the Ht phase. The peak intensity at a $Q$ value of ~0.82 $\text{Å}^{-1}$ (Figure 6b, peak assigned to Ht) appears to grow steadily throughout the period studied, and in fact, its growth is seen to follow a logarithmic function with respect to reaction time, as shown in Figure 7. In contrast, the peak intensity at ~0.78 Å$^{-1}$ (assigned to Hc) stops increasing after an initial period of logarithmic-type growth (i.e., during the initial 15 hours) and then decreases considerably by 5 months (Figure 7). One possible explanation for this decreasing trend at the advanced curing age is the slight increase in carbonation of the Hc phase, leading to changes to the interlayer spacing. As shown in Figure 6b-c, the $Q$ values (at ~0.81, 1.62 and 1.66 Å$^{-1}$) of a slightly more carbonated Hc phase (denoted as Hc', with a chemical composition of $Ca_4Al_2(OH)_{12.4}(CO_3)_{0.8}(H_2O)_4$ [41]) are greater than those of the Hc phase (at ~0.78, 1.55 and 1.62 Å$^{-1}$). Another possible explanation of the decreasing trend of the Hc phase in Figure 7 is a transformation of the Hc phase to a monocarboaluminate phase ($C_4A\underline{c}H_{12}$, abbreviated as Mc) since this transformation has been observed in $Na_2CO_3$-activated GGBSs [72] and the Mc phase has been predicted to be the stable AFm phase to form (as opposed to Hc phase) according to thermodynamic modeling [80]. The XRD data in Figure 6b-c are also generally consistent with this possible transformation of Hc to Mc, specifically since the $Q$ values of the Mc phase at ~0.83 and 1.66 Å$^{-1}$ are greater than those of the Hc phase at ~0.78 and 1.62 Å$^{-1}$. However, due to the evident misalignment of $Q$ values between the XRD data and some of the phases (as seen in Figure 6b-c) it is currently impossible to discern from these data if Hc transforms into Hc' and/or Mc. What is clear from this analysis is that the logarithmic growth identified as the Ht phase in Figure 7 may also contain contributions from the Hc' and/or Mc phase.

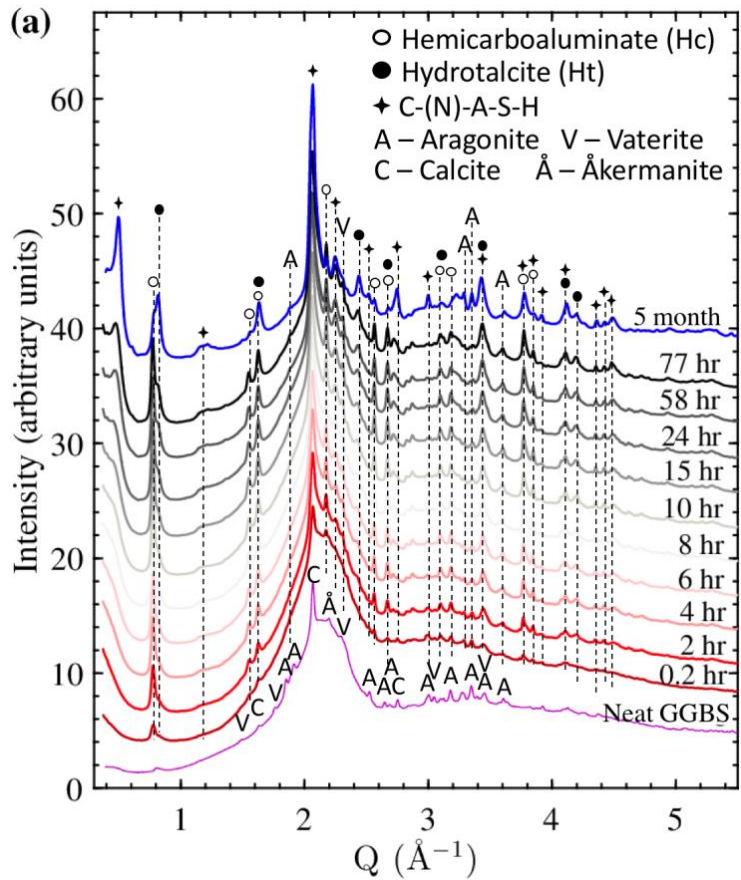

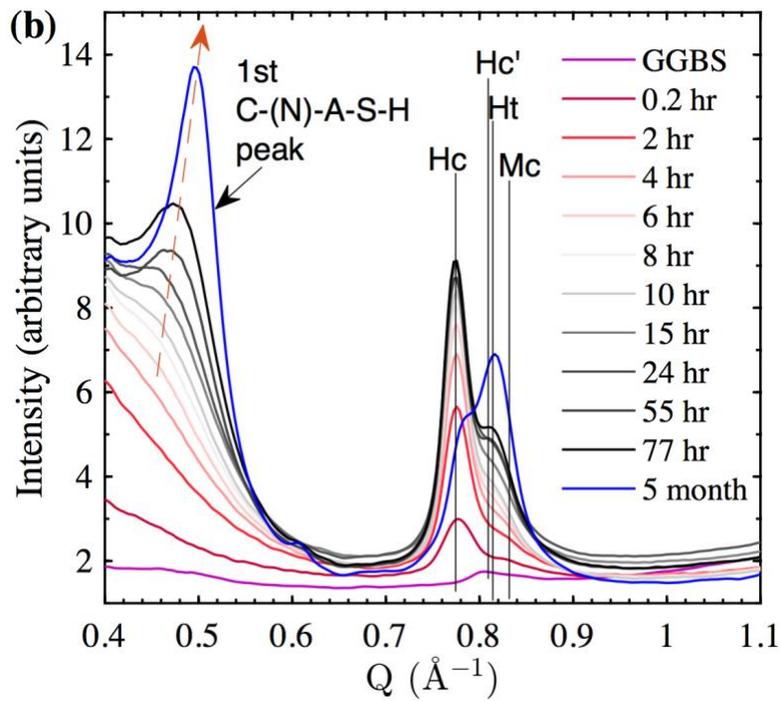

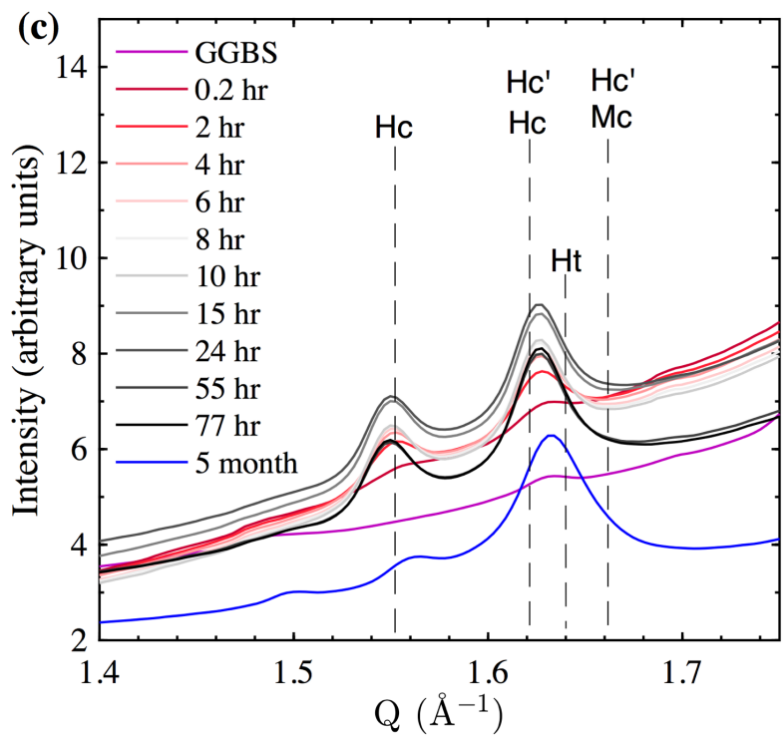

Figure 6. Evolution of the high-resolution synchrotron-based XRD patterns as a function of reaction time over a $Q$ range of (a) 0.3-5.5, (b) 0.4-1.1, and (c) 1.4-1.75 Å$^{-1}$. The XRD patterns in (a) are given as a stacked plot. The peak positions for Ht, Hc, Hc' and Mc phases are based on the structural representations of ICSD #107626 [42], ICSD #263124 [41], ICSD #263123 [41] and ICSD #59327 [81], respectively.

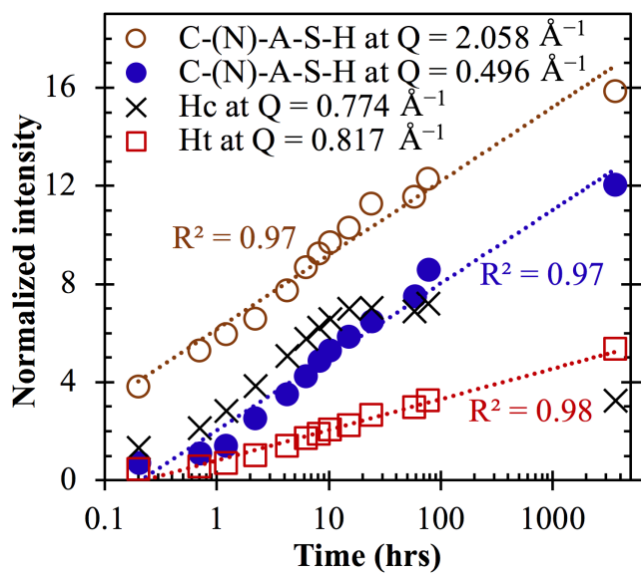

Figure 7. Evolution of XRD peak intensity at different $Q$ values (corresponding to C-(N)-A-S-H gel, Hc and Ht phases) as a function of reaction time. The $R^2$ values in the figure are the goodness of fit of the experimental data obtained using logarithmic functions. Note that the time axis is logarithmic.

Another notable observation in Figure 6a-b is that the basal peak of the C-(N)-A-S-H gel (at a $Q$ value of ~0.45-0.50 Å$^{-1}$) increases in intensity and becomes more defined and narrower after 24 hours of reaction. This increase in intensity appears to be logarithmic with respect to reaction time (Figure 7), and a similar growth pattern is also seen for other characteristic peaks of the C-(N)-A-S-H gel (e.g., the $020/\bar{2}20$ reflection at ~2.06 Å$^{-1}$, Figure 7). The emergence and narrowing of the basal peak indicate an establishment of interlayer ordering, either due to an increased formation of C-(N)-A-S-H gel and/or an atomic rearrangement of gel that has already precipitated. It is also seen in Figure 6b that the basal peak shifts from around ~0.47 Å$^{-1}$ at 58-77 hours to ~0.50 Å$^{-1}$ at 5 months, indicating a reduction in basal spacing from ~13.4 to ~12.6 Å. This reduction in basal spacing for the C-(N)-A-S-H gel could be caused by several factors. First, the formation of inner C-(N)-A-S-H gel product that dominates at the advanced age of curing in alkali-activated GGBS has been shown to exhibit higher Ca/Si than that of outer C-(N)-A-S-H gel formed during the early stage [39], which could lead to smaller basal spacing [67, 70]. Second, it becomes increasingly more difficult to access water at later ages due to the densification of reaction products covering the unreacted GGBS; hence, it is possible that the C-(N)-A-S-H gel formed at later ages has less interlayer water [70]. Third, the generation of crystallization pressure due to product formation in an increasingly restricted space [82] could also lead to a smaller basal spacing, as illustrated for synthetic C-S-H gels under hydrostatic pressure [83]. Nevertheless, there are other factors that could have an impact on the basal spacing such as Al/Si ratio [63, 64, 80] and alkali content [66, 69]. It is also important to keep in mind that changes in scattering in this region may be due to pore structural evolution [32].

### 3.4 Evolution of the local atomic structure during alkaline activation

Figure 8 shows how the X-ray PDF of NaOH-activated GGBS evolves as a function of reaction time up to 5 months. It is clear that most major atom-atom correlations out to ~40 Å undergo significant change during the alkaline activation reaction. The increase in the intensity of atom-atom correlations above ~5 Å is mainly due to the formation of the main binder gel (C-(N)-A-S-H) with some contributions from the secondary phases, as determined from analysis of Figure 4a. The assignment of the atom-atom correlations in Figure 8b is mainly based on the partial X-ray PDFs of the neat GGBS (Figure 2b) and the C-(N)-A-S-H gel in the 5-month sample (Figure 4b), since, as shown in Figure 4a, the secondary LDH phases (i.e., Ht and Hc) do not strongly contribute to the experimental PDF data.

The changes seen in Figure 8b can be grouped according to atom-atom correlation type and behavior. For example, the Ca-O correlations possess a clear trend across the entire $r$ range. The nearest Ca-O correlation at ~2.34 Å increases in intensity and shifts to a larger distance as the reaction progresses. The second nearest Ca-O correlation, initially located at ~4.3 Å due to the presence of a large amount of GGBS, is seen to evolve by the concurrent decrease of the 4.3 Å peak due to GGBS dissolution and emergence of a new peak at ~4.5 Å as C-(N)-A-S-H gel forms. Similarly, the decrease in intensity of the peak at ~7.09 Å and growth of the peak at 7.34 Å are mainly associated with changes occurring to the fourth nearest Ca-O correlation, although this region also contains contributions from Ca-Ca, Ca-Si, Si-Si and O-O correlations (see Figure 4b). These changes to the Ca-O correlations can be attributed to alterations of the local bonding environment of calcium and silicon atoms as GGBS dissolves and C-(N)-A-S-H gel precipitates, and the increase of nanoscale ordering associated with the higher structural coherence of C-(N)-A-S-H gel compared with amorphous GGBS. As an example, the Ca-O-Ca angular distribution and Ca coordination of GGBS and C-(N)-A-S-H gel are displayed in Figures S1 and S2 of the Supplementary Material, which show that there are clear differences in the local bonding environments of these phases. The kinetics of the first, second and fourth nearest Ca-O correlation changes can be elucidated by quantitative analysis of the PDF peak intensity and/or peak position, as shown in Figures 9a and 10a-b, respectively. These figures show that the kinetics of reaction generally follow a logarithmic function with time, where simple linear regressions of the data result in $R^2$ values between 0.93 and 0.99.

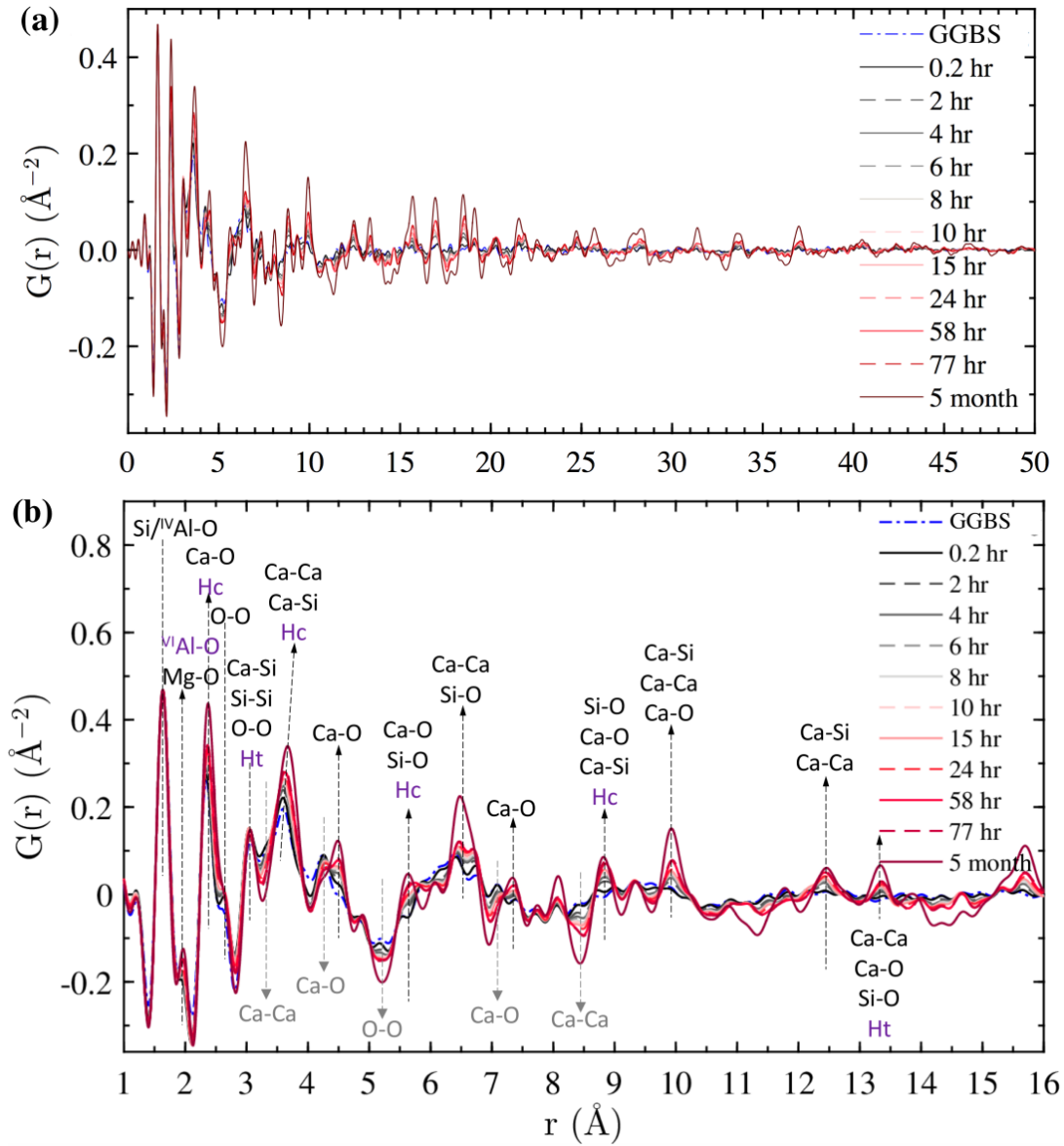

Figure 8. Evolution of the X-ray PDF as a function of reaction time for NaOH-activated GGBS over an *r* range of (a) 0-50 Å and (b) 1-16 Å. The peak assignments in (b) are based on the simulated PDFs of the different phases, with those for the C-(N)-A-S-H gel and secondary phases (i.e., Ht and Hc) labeled in black and purple colors, respectively. The assignments labeled with grey text in (b) have tentatively been assigned to GGBS. Note that only the major atom-atom correlations that likely contribute to the PDF data are shown in (b). Minor contributions have not been included.

Other apparent changes to the atom-atom correlations in Figure 8b include the shoulder at ~3.3 Å attributed to the nearest Ca-Ca correlation in GGBS (Figure 2b) that diminishes as the reaction progresses. Meanwhile, the peak at ~3.6 Å grows continuously with the progress of reaction (Figure 8b) due to the formation of C-(N)-A-S-H gel and associated emergence of the nearest Ca-Ca and Ca-Si correlations and the second nearest Si-O correlation (these correlations were previously discussed in the context of Figures 4b and 5). Again, these time-dependent changes (either increase or decrease) can be approximately described using logarithmic functions, as illustrated in Figure 10c. Similar logarithmic growth and decay of other atom-atom correlations, including those located at ~1.95, ~5.63, and ~8.43 Å, are seen in Figures 9b and 10d, which are tentatively assigned to (i) Mg-O/$^{VI}$Al-O correlation in the LDH phases, (ii) Ca-O, Ca-Ca, Si-O and Si-Si correlations in C-(N)-A-S-H gel and O-O, Ca-O, and Ca-Ca correlations in the Hc phase, and (iii) Ca-Ca correlation in the neat GGBS, respectively.

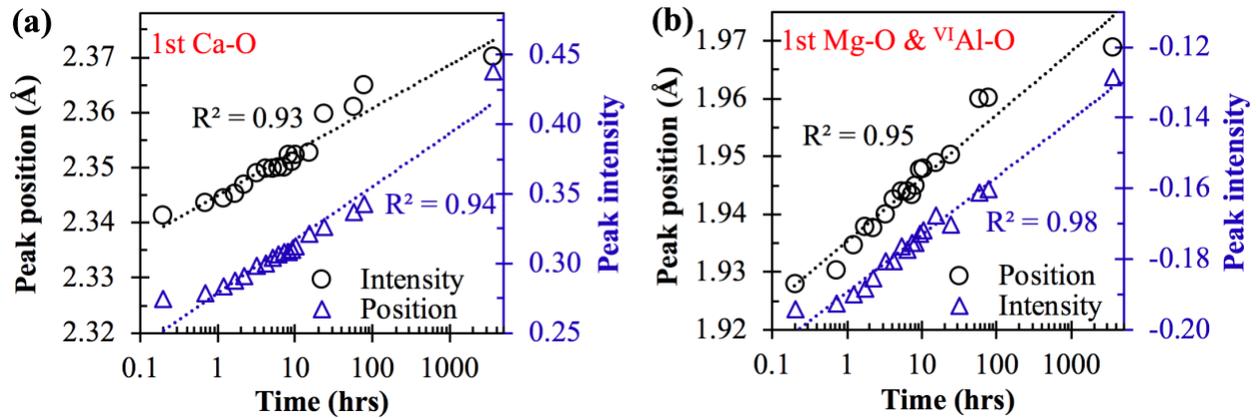

Figure 9. Evolution of the PDF peak position and intensity for (a) first nearest Ca-O and (b) first nearest Mg-O/$^{VI}$Al-O correlations as a function reaction time. For each dataset, the $R^2$ value for the linear regression on a logarithmic time scale is given in the figure.

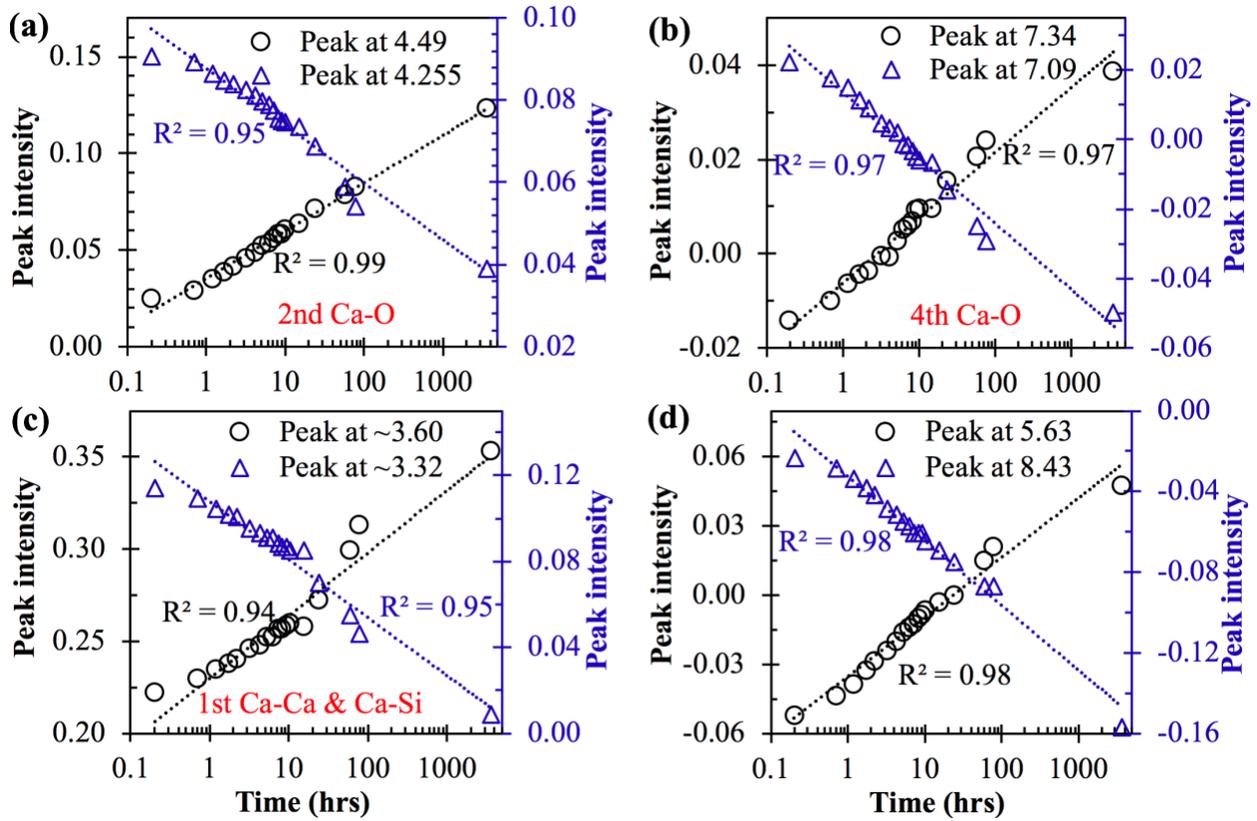

Figure 10. Evolution of PDF peak intensity as a function of reaction time for (a) the second nearest Ca-O correlation (located at ~4.26 Å in GGBS and ~4.49 Å in the C-(N)-A-S-H gel), (b) the fourth nearest Ca-O correlation (located at ~7.09 Å in GGBS and ~7.34 Å in the C-(N)-A-S-H gel), (c) the first nearest Ca-Ca correlation (located at ~3.32 Å in GGBS and ~3.60 Å in the C-(N)-A-S-H gel), and (d) the peaks located at ~8.43 and 5.63 Å. For each dataset, the $R^2$ value for the linear regression on a logarithmic time is given in the figure. Note that the peaks at ~3.60 and 7.34 Å also contain considerable contributions from other atom-atom correlations, as discussed in the text.

## 3.5 Extent of reaction

Quantification of the different phases, i.e., the reaction products and unreacted GGBS, has been performed using the X-ray PDF data collected at each reaction time, similar to the quantification approach used to analyze the 5-month PDF data as described in Section 3.2. At each timestep, the quantification has been carried out based on four possible phases: unreacted GGBS, C-(N)-A-S-H gel, Ht, and Hc. The resulting scale factors of these phases have been used to obtain the relative quantity of each phase, $q_i$, estimated at each measurement time, $t$, using the following equation (Equation (4)):

$$q_{i_t} = \left(\frac{SF_i}{\sum SF_i}\right)_t \quad (4)$$

where $SF_i$ is the scale factor of the $i^{th}$ phase at time $t$.

Based on the relative proportion of each phase, the degree of reaction (DOR) of the GGBS at time $t$ can be estimated using the following equation (Equation (5)):

$$DOR_t = 1 - q_{GGBS_t} \quad (5)$$

where $q_{GGBS_t}$ is the quantity of unreacted GGBS in the paste at time $t$, calculated using Equation (4).

Figure 11a displays the evolution of the quantities of the different phases as the alkaline activation reaction progresses, clearly showing that both the dissolution of GGBS and the formation of C-(N)-A-S-H gel follow a logarithmic-type behavior, in agreement with the XRD results in Section 3.3 (Figure 7) and the PDF peak intensity/position results in Section 3.4 (Figures 9 and 10). The formation of the secondary Ht phase is also seen to have a logarithmic-type behavior with time, which is consistent with the XRD data in Figure 7. However, the trend for Hc phase formation seen in Figure 11a (continual growth up to 5 months) is different from that obtained from the XRD data (Figure 7) where Hc phase stopped forming at ~10 hours and its total amount decreased between 77 hours and 5 months. To attempt to reconcile these differences we have also carried out the PDF quantification approach with the inclusion of an Mc or Hc' phase (in addition to unreacted GGBS, C-(N)-A-S-H gel, Ht and Hc phases), since we postulated in Section 3.3 that a transformation of Hc to the Mc or Hc' phase may have occurred at the advanced age of curing. The results, seen in Figure S3a-d of the Supplementary Material, do not show such a transformation of Hc to the Mc or Hc' phase and instead show, once again, a logarithmic-type growth for the Hc phase. This discrepancy between the XRD and PDF results regarding the trend of growth of the Hc phase may be caused by the limitations associated with the PDF quantification method used in this investigation, as discussed in Section 6 of the Supplementary Material.

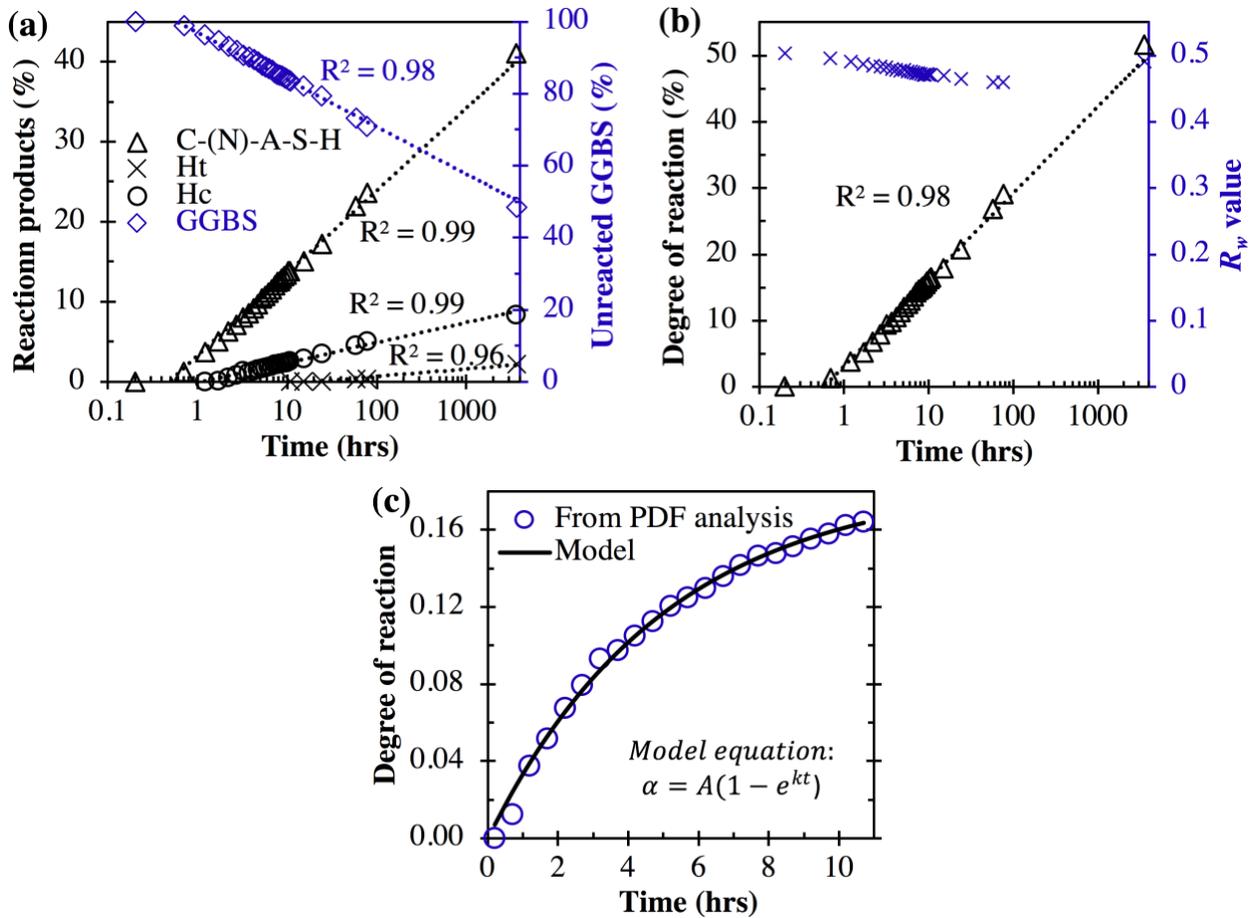

Figure 11. Evolution of (a) the relative quantities of the different phases and (b) the degree of reaction (DOR) in the NaOH-activated GGBS (in %) as a function of reaction time, obtained using the PDF quantification method. Also shown in (b) are the $R_w$ values achieved for the least squares refinement process at each timestep. A linear regression of each dataset using a logarithmic time scale is shown in the figure along with the goodness of fit ($R^2$ value). (c) The fit of the DOR data during the first ~11 hours with a modified pseudo-single step first-order reaction model ($\alpha = A(1 - e^{kt})$).

The DOR data (in Figure 11b) obtained using the PDF quantification method agree reasonably well with results from other quantification methods that have been applied to similar types of alkali-activated GGBS pastes. For instance, the DOR for our 5-month NaOH-activated GGBS sample is estimated to be ~50%, which is consistent with values reported for similar hydroxide-activated GGBS pastes cured for 100-180 days (~40-60%) obtained using NMR [14, 15] and backscattered SEM image analysis [14, 17, 18]. Note that the $R_w$ values achieved for the least squares refinement are in the range of 0.47-0.50, which are comparable with several previous investigations on the modeling of the atomic structure of amorphous GGBS (0.35-0.47) [50, 61], iron-rich slag (0.38) [84], magnesium carbonate (0.48) [85] and metakaolin (0.77) [86].

We attempted to use simple rate equations (1$^{st}$, 2$^{nd,}$ and 3$^{rd}$ order) to fit the data in Figure 11b, however, none of these rate equations adequately describe the reaction kinetics for NaOH-activated GGBS. This is consistent with a previous investigation on the reaction kinetics of different alkali-activated systems [30], indicating the complexity of the alkali-activation process. Nevertheless, an altered pseudo-single step first-order rate equation was shown in ref. [30] to give a reasonable description of the reaction kinetics for several alkali-activated materials (including a NaOH-activated GGBS) during the early stages of reaction. Here, we evaluated whether a similarly modified pseudo-single step first-order rate equation in the form of $\alpha = A(1 - e^{kt})$ can be used to describe the reaction kinetics of NaOH-activated GGBS during the initial ~11 hours. The fit result is present in Figure 11c, which shows that the early stages of reaction are accurately captured by this rate equation (with a weighted sum of the squared residue $\chi^2$ of ~0.0003). The corresponding rate constant $k$ for the fit is 0.21 h$^{-1}$, which is higher than that obtained in ref. [30] for a NaOH-activated GGBS (0.04 h$^{-1}$), where data in the range of 2-10 hours were used to fit the function $\alpha = c + 1 - e^{kt}$. However, if $\alpha = A(1 - e^{kt})$ is used to fit the same data from ref. [30] over the range of 0-10 hours (see Figure S5 in the Supplementary Material for details) a slightly lower level of agreement is achieved, with a $\chi^2$ of ~0.006. The corresponding rate constant $k$ for the fit in Figure S5 is ~0.20 h$^{-1}$, which is almost the same as the fit results from Figure 11c, although the GGBS composition, activator concentration and method of calculating DOR are different.

The PDF phase quantification method introduced here provides an alternative approach for quantifying the amorphous phase(s) present in a multi-phase amorphous/disordered/crystalline system, in addition to the well-known PONKCS method based on reciprocal space XRD data. One advantage of the current approach that is based on analysis of scattering data in real space is that it does not require the use of an internal or external standard for the quantification process (as often needed for the PONKCS method). Furthermore, real space X-ray PDF data provide valuable information on the local atomic structure of the amorphous phase(s) (such as GGBS); information that is hard to obtain from the same data if given in reciprocal space. Another advantage of the method presented here, compared with other experimental methods that have also been used to quantify the DOR of amorphous aluminosilicates in the cements field (including NMR [10, 14-16], backscattered SEM [8, 17-19], selective dissolution [8, 9, 19-21], and TGA [9, 10, 18, 22, 23], as has been briefly outlined in the Introduction) is that the use of synchrotron-based X-rays allows for fast acquisition (i.e., seconds to a few minutes) of high-resolution data enabling phase quantification of *in situ* reactions involving cement-based systems without the need to artificially "freeze" the reaction prior to measurement. A drawback of this method is that it does need accurate structural representations for all major phases involved during the chemical reaction, including the amorphous GGBS phase, which limits quantification to systems where structural representations of all phases are available.

### 3.6 Comparison of the PDF quantification data with other experimental data

As outlined in the Introduction, there are many other *in situ* experimental techniques that have been used to investigate the reaction kinetics of cement-based systems. Hence, to further validate the PDF quantification method, we have collected (or obtained from the literature) complementary *in situ* experimental data on the formation of NaOH-activated GGBS for comparison. The complementary data include (i) ICC and FTIR for the same NaOH-activated GGBS binder, and (ii) QENS data from ref. [44] for a NaOH-activated GGBS based on different GGBS sources and $Na_2O$ content. Below the complementary reaction kinetics data from ICC and FTIR are compared with the DOR data obtained from the PDF analysis. Refer to the Supplementary Material for comparison of the PDF-derived DOR data with QENS. In summary, the quantification data based on PDF analysis are consistent with all three complementary techniques.

*3.6.1 Isothermal conduction calorimetry data*

ICC data collected on the same NaOH-activated GGBS investigated using the PDF quantification method are presented in Figure 12a, which show that the ICC heat flow curve exhibits two peaks: one major peak at the start of the measurement and second less intense peak at ~1.5 hours, both of which are similar to our previous ICC data on hydroxide-activated GGBS [44]. The initial major peak is attributed predominantly to the wetting of particle surfaces, while the second peak is mainly due to the formation of reaction products [44]. If we compare the ICC data after 0.2 hours of reaction with the PDF-derived DOR data from Figure 11b (see Figure 12b for comparison) we clearly see that the two datasets are aligned. The smaller slope during the initial hour of the ICC data (compared to the slope for the data after 1 hour) has been captured by the PDF-derived DOR data. To further assess the validity of the PDF-based quantification method the normalized ICC data at each timestep have been plotted against the PDF intensity values of specific atom-atom correlations, as shown in Figure S6 of the Supplementary Material. This figure reveals that there are clear linear trends between individual atom-atom correlations and the cumulative heat.

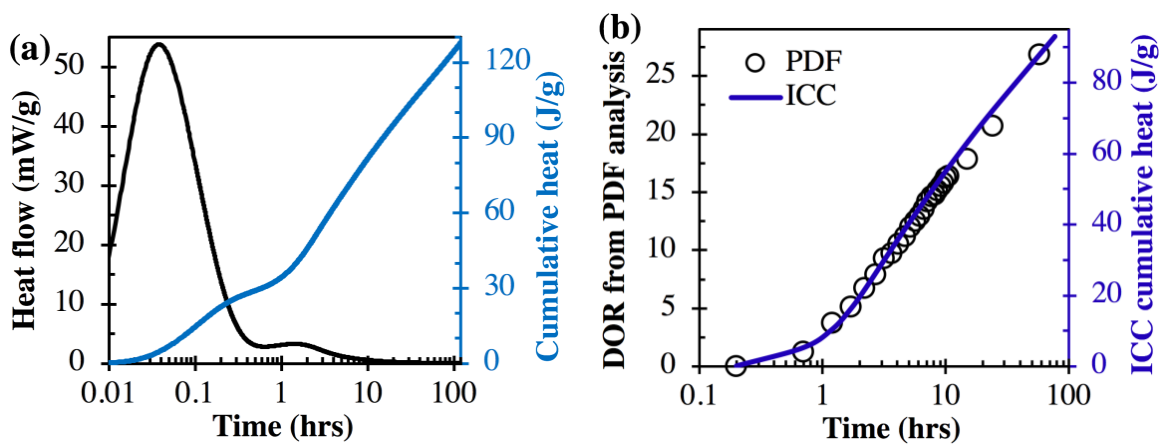

Figure 12. (a) ICC data, and (b) comparison of the ICC cumulative heat with the degree of reaction (DOR) from PDF analysis as a function of reaction time. The ICC data in (b) have been normalized by assuming that the cumulative heat is zero at 0.2 hours (i.e., the heat attributed to wetting of particle surfaces has been removed).

*3.6.2 Fourier transform infrared spectroscopy data*

Figure 13a shows the evolution of the main asymmetric Si−O−T stretching band of the FTIR data collected on the same NaOH-activated GGBS binder over a period of 8 days. During the initial hour the spectrum remains essentially unchanged while at ~2 hours a feature emerges at ~935 cm$^{-1}$ indicating gel formation (feature is typical of C-S-H-type gels and tobermorite structures [87-89]). After ~4-8 hours other characteristic features of C-S-H-type gels (e.g., ~805, ~895, and ~1000 cm$^{-1}$) become visible and grow with time. These observations, and specifically the emergence of peaks attributed to the main binder gel, are consistent with our previous investigation on a similar type of hydroxide-activated GGBS [44]. Based on the literature [44, 87-89], the main band at ~935 cm$^{-1}$ and the two shoulders at ~895 and ~1000 cm$^{-1}$ are assigned to asymmetric Si−O−T stretching of $Q^2$ units with different local environments. The band at ~805 cm$^{-1}$ is assigned to Si−O stretching of $Q^1$ tetrahedra [89].

It is possible to estimate the DOR for the hydroxide-activated GGBS by deconvolution of the main FTIR band (at 600-1100 cm$^{-1}$) using multiple Gaussians and an unreacted GGBS component (Figure 13b), as outlined in our previous investigation [44]. An example deconvolution is shown in Figure 13b, where the shoulder region at ~840-860 cm$^{-1}$ in the FTIR spectra of NaOH-activated GGBS has been attributed to unreacted GGBS since this feature does not seem to grow throughout the period investigated while the features next to it (i.e., ~805 and ~895 cm$^{-1}$) emerge and grow [44]. Hence, the contribution from the unreacted GGBS to the overall band can be determined by scaling the $v_{as}$ Si—O—T band of the neat GGBS such that the spectrum at ~840-860 cm$^{-1}$ overlap with that of NaOH-activated GGBS, as highlighted by the red dashed circle in Figure 13b. This is based on the assumption that overall shape of $v_{as}$ Si—O—T band of unreacted GGBS remains the same as the neat GGBS. Based on the deconvolution of the FTIR spectrum (Figure 13b), we have estimated the relative amount of each individual deconvoluted band, $r_i$, and the overall DOR, for each timestep using the equations below:

$$r_i = \frac{A_i}{A_{GGBS}+\sum A_i} \tag{6}$$

$$DOR = \frac{\sum A_i}{A_{GGBS}+\sum A_i} \tag{7}$$

where $A_i$ is the area of the $i^{th}$ Gaussian and $A_{GGBS}$ is the area of the unreacted GGBS in Figure 13b. The evolution of DOR from the FTIR analysis is compared with that obtained from PDF analysis in Figure 13c, which shows that the overall trend and the DOR values are in general agreement but the DOR from FTIR contains more scatter. We have also compared the evolution of the individual deconvoluted bands (i.e., $r_i$ in Equation (6)) with the DOR from the PDF analysis in Figure S7 of the Supplementary Material, which reveals similar trends of evolution for all the individual deconvoluted bands, i.e., logarithmic functions of reaction time with $R^2$ values of 0.70-0.92. This suggests that all four bands (~805, ~895, ~935 and ~1000 cm$^{-1}$) are associated with the formation of C-(N)-A-S-H gel. A major difference between the two methods (i.e., FTIR and PDF to obtain the DOR data) is that the FTIR method does not capture the formation of C-(N)-A-S-H gel during the initial hour (see Figure 13c) in contrast with the PDF method where gel formation during the initial hour is clearly evident (as seen in Figure 11a). An interesting, yet subtle observation from Figure 13a is that the main asymmetric Si—O—T stretching band position shifts progressively to larger wavenumbers as a function of reaction time (the shift is clearly evident in Figure S8 of the Supplementary Material by the evolution of the main deconvoluted band at ~935

cm$^{-1}$). This shift suggests that the gel forming during this period (192 hours) becomes increasingly more polymerized with the progress of reaction.

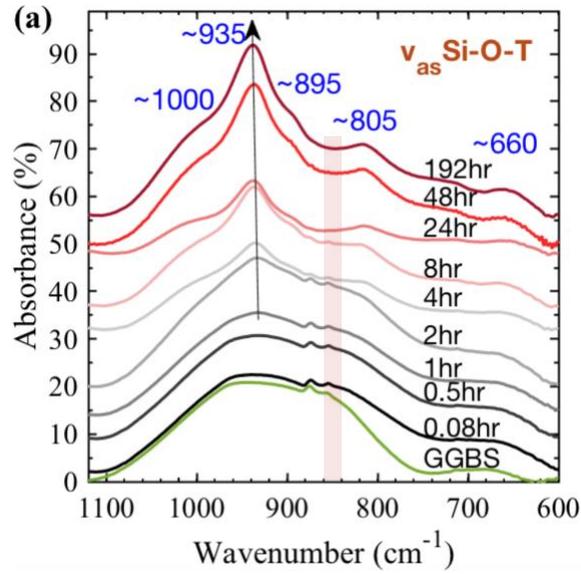

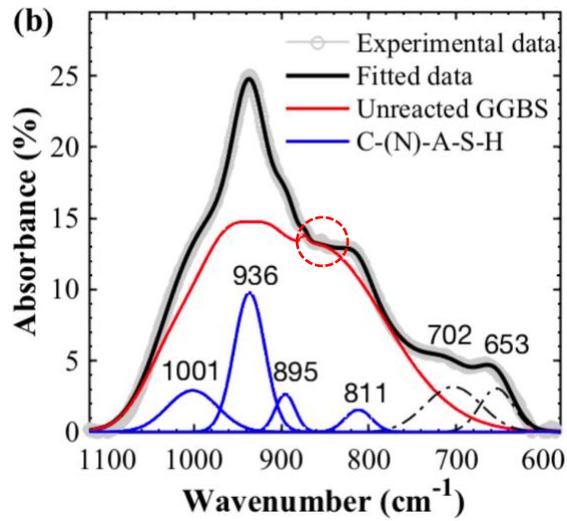

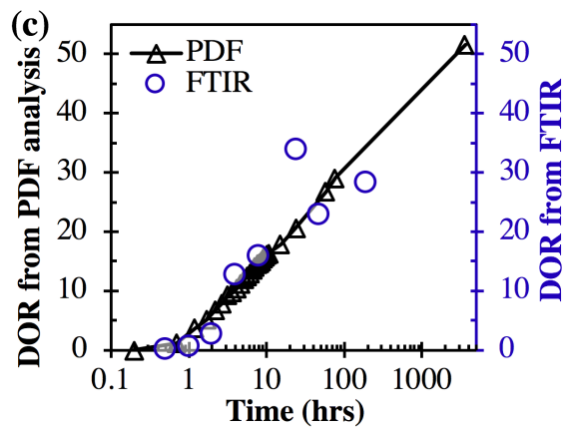

Figure 13. (a) Evolution of the FTIR spectra (from 600-1100 cm$^{-1}$) as a function of reaction time collected on NaOH-activated GGBS. (b) The deconvolution of a typical FTIR spectrum (8 hour data) using Gaussians to model the C-(N)-A-S-H gel and the FTIR spectrum of neat GGBS to account for unreacted GGBS. (c) Comparison of the evolution of the degree of reaction (DOR) values obtained using PDF and FTIR analysis.

## 3.7  Broader impact & limitations

### 3.7.1  Broader Impact

The quantification method outlined in this investigation that combines atomistic modeling of disordered phases and/or structural representations obtained from the literature with *in situ* X-ray PDF analysis is seen to accurately capture the quantity of amorphous/disordered and crystalline phases throughout the reaction process of NaOH-activated slag. Moreover, the use of structural representations for the quantification method, as opposed to solely experimental datasets, allows for changes in the local bonding environments to be identified and analyzed, providing additional insight on structural changes that occur during the reaction. In general, such an approach can be applied to other complex chemical processes that involve transformations between amorphous, disordered, and crystalline phases. These processes include those involving other AAMs, blended cements, bioglass dissolution, mineral dissolution, glass corrosion, and nuclear waste encapsulation [1-4]. The use of *in situ* synchrotron X-rays also enables these chemical processes to be probed in an undisturbed state with high temporal resolution. This is particularly important for chemical processes involving phase formation/transformation processes on the seconds-to-minutes time scale, as is the case for formation of cements. Lastly, as exemplified here, the quantification of the different phases and the DOR as a function of time can be combined with reaction kinetic models to provide additional insight on the type of processes that are occurring during the *in situ* reaction.

### 3.7.2  Limitations

Although the method introduced here is seen to give an accurate quantitative description of GGBS dissolution, phase formation and DOR in the NaOH-activated GGBS (generally consistent with the ICC and FTIR data, as well as BWI data from QENS analysis), several limitations need to be considered and discussed. As alluded to in Section 3.5, as it stands this method is less sensitive to the evolution of minor phases and hence may not be able to capture all the potential transformations occurring during the alkaline activation process; for example, the possible transformation of Hc to the Mc or Hc' phase at the later stages of reaction in NaOH-activated GGBS, as indicated by the *in situ* XRD data (Figures 6 and 7). Furthermore, the application of this method requires accurate structural representations for all the potential individual phases in the system at any given time, yet the generation of accurate structural representations can be particularly challenging for amorphous materials. In this article, we have generated a reasonable atomic structural representation for the amorphous GGBS by using force-field MD simulations. However, there are still obvious discrepancies between the experimental and simulated (from MD simulations) PDF data, as illustrated in Figure 2. An improved structural representation for the amorphous phase as well as the reaction product phases may further enhance the performance of this quantification method.

Possible future work includes optimization of the force-field MD-generated structural representation of GGBS using DFT calculations to obtain a more accurate structure. However, this would require the use of a smaller system size (i.e., fewer atoms) such that the DFT calculations are not computational prohibitive, which may make it difficult to accurately describe the structural arrangements within an amorphous phase. Alternatively, the development of more accurate force-fields for modeling GGBS will be extremely valuable, especially considering the need to cover an extremely large compositional range of glasses relevant to AAM and blended cement technologies in order to establish important composition-structure-reactivity relationships. Furthermore, although the C-(N)-A-S-H gel structural representation used here has been optimized using DFT calculation, several important features of C-(N)-A-S-H gels in NaOH-activated GGBS have not been replicated by the structural representation. This includes (i) the disordered nature of the gel (treated as crystalline in the simulation), (ii) changing Ca/Si ratio in the gel during the alkali-activation reaction (a constant ratio associated with the crystalline representation was used in this investigation), and (iii) finite silicate chain length (the structural representation had an infinite chain length).

## 4  Conclusions

Chemical processes involving amorphous-to-disordered/crystalline transformations (and vice versa) are ubiquitous in many important natural and engineering applications, including sustainable cements such as alkali-activated materials (AAMs). However, our ability to (i) capture changes in both phase assemblage and local bonding environments and (ii) quantify the rate and extent of such changes is somewhat limited, particularly for complex systems that also contain multiple amorphous and disordered components. In this study, the phase evolution and atomic structural changes occurring during the formation of a sodium hydroxide-activated ground granulated blast-furnace slag (GGBS) have been investigated using *in situ* high-resolution X-ray diffraction (XRD) and X-ray pair distribution function (PDF) analysis. The XRD results showed that the reaction products, including the main sodium-containing calcium-alumino-silicate-hydrate (C-(N)-A-S-H) gel phase and two layered double hydroxide (LDH) secondary phases (i.e., calcium hemicarboaluminate (Hc) and a hydrotalcite-like phase (Ht)) form immediately after mixing. The C-(N)-A-S-H gel and the Ht phase grow continuously with the progress of reaction. Analysis of the gel basal peak suggested an increase in the interlayer ordering and reduction of basal spacing for the gel as the reaction evolved. *In situ* X-ray PDF analysis revealed a continuous increase of structural ordering attributed to the loss of amorphous GGBS and the emergence of a more ordered C-(N)-A-S-H gel along with contributions from the crystalline secondary phases.

Based on the *in situ* real space PDF data, we outlined a method for quantification of the amorphous-to-disordered/crystalline transformations that occur during sodium hydroxide activation of GGBS. First, we generated a detailed atomic structural representation for the amorphous GGBS by use of a melt-and-quench approach utilizing force-field molecular dynamics (MD) simulations, which was seen to be in agreement with the X-ray PDF data. Second, the evolution of the individual phases was quantified as a function of reaction time, specifically by comparing the *in situ* experimental PDF data of the formation of NaOH-activated GGBS with simulated PDF data obtained from the MD-generated structural representation of GGBS together with literature-derived structural representations of the reaction products. These structural representations also enabled, for the first time, the assignment of major atom-atom correlations to the individual PDF peaks seen in the *in situ* experimental data out to ~14 Å. A major outcome of the PDF phase quantification of NaOH-activated GGBS was calculation of the overall degree of reaction (DOR), which was seen to agree with existing literature data. Moreover, the evolution of the DOR during the early stages of reaction (~0.2-11 hours) was seen to be accurately captured by a modified pseudo-single step first-order reaction model ($\alpha = A(1 - e^{kt})$). However, if the later stages of reaction were included (up to 5 months), the evolution of DOR was more accurately described by a logarithmic function. Validation of the PDF-derived DOR was carried out via comparison with isothermal conduction calorimetry (ICC) and Fourier transform infrared spectroscopy (FTIR) data collected on the same NaOH-activated GGBS and *in situ* quasi-elastic neutron scattering (QENS) data collected on a different NaOH-activated GGBS, as reported in the literature. Hence, this investigation has highlighted the power of combining atomistic modeling with *in situ* X-ray PDF analysis for studying the reaction mechanisms and kinetics occurring in a complex material system involving amorphous-to-disordered/crystalline transformations.

# 5 Supplementary Material

Estimation of the CMAS glass density at different temperatures; Calculation of the PDF and partial PDFs from a structural representation; Ca-O-Ca angular distribution for GGBS and C-(N)-A-S-H gel structural representations; Ca coordination number of GGBS and C-(N)-A-S-H gel structural representations; PDF phase quantification based on five phases; Impact of zeroing negative scale factors on the phase quantification results; Analysis of the reaction data for a NaOH-activated GGBS from White et al., 2013; Comparison of PDF and ICC data; Analysis of the individual deconvoluted FTIR bands; Calculation of the relative quantity of chemically bound H-atoms based on phase quantification using PDF data.

# 6 Acknowledgments

This work was supported by the National Science Foundation under Grant No. 1362039 and ARPA-E under Grant No. DE-AR0001145. Part of KG's participation was enabled by a Charlotte Elizabeth Procter Fellowship from Princeton University. The authors would like to acknowledge the use of the 11-ID-B beamline at the Advanced Photon Source, an Office of Science User Facility operated for the U.S. DOE Office of Science by Argonne National Laboratory, under U.S. DOE Contract No. DE-AC02-06CH11357. The authors would like to thank the beamline staff Mr. Kevin Beyer and Dr. Olaf Borkiewicz, and colleagues Dr. Kengran Yang and Dr. Arne Peys for their assistance during the synchrotron experiment. The authors also acknowledge the use of computational resources supported by Princeton Institute for Computational Science and Engineering (PICSciE) and the High Performance Computing Research Center at Princeton University.

# 7 References

[1] G.S. Frankel, J.D. Vienna, J. Lian, J.R. Scully, S. Gin, J.V. Ryan, J. Wang, S.H. Kim, W. Windl, J. Du, A comparative review of the aqueous corrosion of glasses, crystalline ceramics, and metals, NPJ Mater. Degrad., 2 (2018) 15.
[2] J. Du, J.M. Rimsza, Atomistic computer simulations of water interactions and dissolution of inorganic glasses, NPJ Mater. Degrad., 1 (2017) 16.
[3] A.E. Morandeau, C.E. White, The role of magnesium-stabilized amorphous calcium carbonate in mitigating the extent of carbonation in alkali-activated slag, Chem. Mater., 27 (2015) 6625-6634.
[4] J. Skibsted, R. Snellings, Reactivity of supplementary cementitious materials (SCMs) in cement blends, Cem. Concr. Res., 124 (2019) 1-16.
[5] Z. Sun, A. Vollpracht, Isothermal calorimetry and *in-situ* XRD study of the NaOH activated fly ash, metakaolin and slag, Cem. Concr. Res., 103 (2018) 110-122.
[6] R.P. Williams, R.D. Hart, A. Van Riessen, Quantification of the extent of reaction of metakaolin-based geopolymers using X-ray diffraction, scanning electron microscopy, and energy-dispersive spectroscopy, J. Am. Ceram. Soc., 94 (2011) 2663-2670.
[7] R. Snellings, A. Salze, K.L. Scrivener, Use of X-ray diffraction to quantify amorphous supplementary cementitious materials in anhydrous and hydrated blended cements, Cem. Concr. Res., 64 (2014) 89-98.
[8] V. Kocaba, E. Gallucci, K.L. Scrivener, Methods for determination of degree of reaction of slag in blended cement pastes, Cem. Concr. Res., 42 (2012) 511-525.
[9] M. Mejdi, W. Wilson, M. Saillio, T. Chaussadent, L. Divet, A. Tagnit-Hamou, Quantifying glass powder reaction in blended-cement pastes with the Rietveld-PONKCS method, Cem. Concr. Res., 130 (2020) 105999.
[10] S. Kucharczyk, M. Zajac, C. Stabler, R.M. Thomsen, M. Ben Haha, J. Skibsted, J. Deja, Structure and reactivity of synthetic CaO-$Al_2O_3$-$SiO_2$ glasses, Cem. Concr. Res., 120 (2019) 77-91.
[11] A. Cuesta, I. Santacruz, S.G. Sanfélix, F. Fauth, M.A.G. Aranda, A.G. De la Torre, Hydration of $C_4AF$ in the presence of other phases: a synchrotron X-ray powder diffraction study, Constr. Build. Mater., 101 (2015) 818-827.
[12] G.V.P. Bhagath Singh, K.V.L. Subramaniam, Method for direct determination of glassy phase dissolution in hydrating fly ash-cement system using X-ray diffraction, J. Am. Ceram. Soc., 100 (2017) 403-412.
[13] K.C. Reddy, K.V.L. Subramaniam, Quantitative phase analysis of slag hydrating in an alkaline environment, J. Appl. Cryst., 53 (2020) 424-434.